\documentclass[a4paper,11pt]{article}
\pdfoutput=1 

\usepackage{jinstpub} 
\graphicspath{{./img/}}

\usepackage[acronym,nomain]{glossaries}

\usepackage{xspace}

\usepackage{comment}

\usepackage{amsmath,psfrag}
\usepackage{float}

\newcommand{\fom}{\mathrm{FOM}}
\newcommand{\fomlf}{\mathrm{FOM}_\mathrm{LF}}
\newcommand{\enoblf}{\mathrm{ENOB}_\mathrm{LF}}

\newcommand{\fomunit}{fJ$/$\emph{conv.-step}}

\title{Ultra-low power 10-bit 50-90\,MSps SAR ADCs in 65~nm CMOS for multi-channel ASICs}

\author{Miros\l{}aw Firlej,}
\author{Tomasz Fiutowski,}
\author[1]{Marek Idzik,\note{Corresponding author.}}
\author{Jakub Moro\'n,}
\author{Krzysztof \'Swientek}

\affiliation{AGH University of Krakow, Faculty of Physics and Applied Computer Science\\Al. Mickiewicza 30, 30-059 Krak\'ow, Poland}

\emailAdd{idzik@fis.agh.edu.pl}

\abstract{The design and measurement results of ultra-low power, fast 10-bit \gls{sar} \gls{adc} prototypes in 65~nm CMOS technology are presented. 
Eight prototype \glspl{adc} were designed using two different switching schemes of capacitive \glspl{dac}, based on MIM or MOM capacitors, and controlled by standard or low-power \gls{sar} logic. The layout of each \gls{adc} prototype is drawn in 60~um pitch to make it ready for multi-channel implementation. A series of measurements have been made confirming that all prototypes are fully functional, and six of them achieve very good quantitative performance.  Five out of eight \glspl{adc} show both integral (INL) and differential (DNL) nonlinearity errors below 1~LSB. In dynamic measurements performed at 0.1~Nyquist input frequency, the effective number of bits (ENOB) between 8.9--9.3 was obtained for different \gls{adc} prototypes. Standard \gls{adc} versions work up to 80--90~MSps with ENOB between 8.9--9.2 bits at the highest sampling rate, while the low-power versions work up to above 50~MSps with ENOB around 9.3 bits at 40~MSps. The power consumption is linear with the sample rate and at 40~MSps it is around 400~$\upmu$W for the low-power \glspl{adc} and just over 500~$\upmu$W for the standard \glspl{adc}. At 80~MSps the standard \glspl{adc} consume about 1~mW.
}
\keywords{ Front-end electronics for detector readout, Analogue electronic circuits, Digital electronic circuits, VLSI circuits}

\begin{document}
\glsdisablehyper 

\newacronym{adc}{ADC}{Analog-to-Digital Converter}
\newacronym{asic}{ASIC}{Application-Specific Integrated Circuit}
\newacronym{dac}{DAC}{Digital-to-Analog Converter}

\newacronym{dsp}{DSP}{Digital Signal Processing}
\newacronym{tdc}{TDC}{Time-to-Digital Converter}

\newacronym{lhc}{LHC}{Large Hadron Collider}
\newacronym{hep}{HEP}{High Energy Physics}

\newacronym{dnl}{DNL}{Differential Non-Linearity}
\newacronym{inl}{INL}{Integral Non-Linearity}

\newacronym{thd}{THD}{Total Harmonic Distortion}
\newacronym{sfdr}{SFDR}{Spurious-Free Dynamic Range}
\newacronym{snhr}{SNHR}{Signal-to-Non Harmonic Ratio}
\newacronym{sinad}{SINAD}{Signal-to-Noise-and-Distortion ratio}
\newacronym{enob}{ENOB}{Effective Number of Bits}
\newacronym{fom}{FOM}{Figure of Merit}

\newacronym{lvds}{LVDS}{Low-Voltage Differential Signalling}
\newacronym{fpga}{FPGA}{Field Programmable Gate Array}
\newacronym{sar}{SAR}{Successive Approximation Register}

\maketitle
\flushbottom
\glsresetall  

\section{Introduction}

In modern and newly designed particle physics detection systems, there is a growing demand for detectors with ever-increasing speed, high granularity, and high channel density. The key part of such a detector is a dedicated multi-channel readout \gls{asic}, which has gained increasing functionality in recent years, slowly becoming a System on Chip (SoC). In particular, the speed of signal processing is increasing; each channel is required to measure the amplitude or time (or both) and convert the result to a digital form. As the number of bits of information increases, so does the demand for faster data transmission. With the above requirements and increasing channel density, ultra-low power consumption per channel is a must. 

A fast, ultra-low power, area-efficient \gls{adc} is one of the indispensable components of a SoC-type readout \gls{asic}. An ultra-low power \gls{adc} with a sampling rate of 40~MSps or more, medium-high resolution, and small pitch is required for multi-channel readout \glspl{asic} in modern and future LHC or other experiments.
Recent developments of such complex readout \glspl{asic} are a 128-channel SALT \gls{asic} for the LHCb Upstream Tracker, which contains an analogue front-end and a 6-bit 40 MSps \gls{adc} in each channel~\cite{SALT}, or a 72-channel HGCROC \gls{asic} for the CMS High Granularity Calorimeter, which contains an analogue front-end, a 10-bit 40 MSps \gls{adc} and a precision TDC in each channel~\cite{HGCROC3}. 
In fact, a fast 10-bit \gls{adc} is one of the most requested and used blocks in the readout of various detector systems~\cite{HGCROC3, SAMPA_2020, TOFHIR, VMM3a, HKROC}.
These and other readout \glspl{asic} for LHC and other experiments have been developed in the 130~nm CMOS process, which has been studied in the past and selected for use in \gls{hep} experiments many years ago due to its very good performance and good radiation tolerance~\cite{radhard_cmos130}. 

For medium- and long-term future experiments, newer CMOS processes will be used, not only because of the higher speed, density, and lower power, but also because of the limited availability in time of current technologies. One of such technology that has already been verified, also in terms of radiation hardness, is the 65~nm CMOS process~\cite{radhard_cmos65}. Several developments of complex readout \glspl{asic} in CMOS 65~nm have already started~\cite{rd53_itk_65, timepix4, lpGBT} and this process will be dominant for the next 5--10 years until a newer one, probably CMOS 28~nm, will take place. For the highest density \glspl{asic}, e.g. pixel detectors, the transition to 28~nm CMOS will be much faster.

The aim of this work is to develop a fast, ultra-low power \gls{adc} in CMOS 65~nm, ready for integration into a multi-channel readout \glspl{asic} for future experiments. The main goals for \gls{adc} are: a sampling rate of at least 40~MSps (but possibly significantly higher), 10-bit resolution, ultra-low power consumption of around 500~$\upmu$W at 40~MSps, small pitch per channel below 100~$\upmu$m, and easy implementation in a multi-channel readout \gls{asic}. 

\section{ADC design}

\begin{figure}[tbp]
 \centering
 \includegraphics[width=0.9\textwidth]{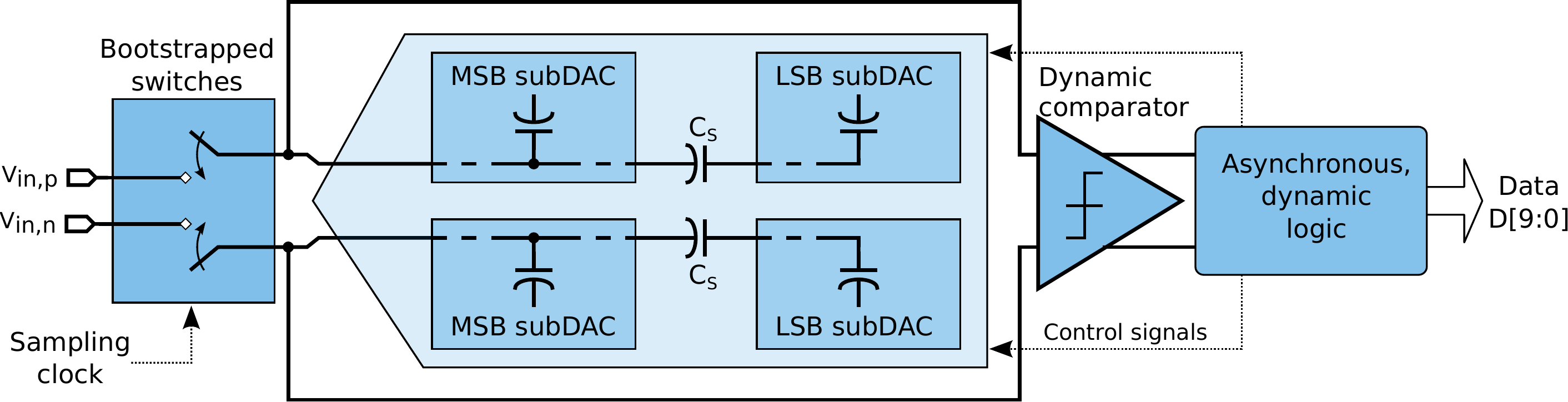}
 \caption{Block diagram of a fully differential 10-bit \gls{sar} \gls{adc} with split \gls{dac} architecture.}
 \label{fig:sar_diagram}
\end{figure}

The demand for ultra-low power \gls{adc} naturally leads to a \gls{sar} architecture with a capacitive \gls{dac}, shown in the block diagram in figure~\ref{fig:sar_diagram}. 
A fully differential \gls{adc} architecture was chosen, comprising a pair of bootstrapped switches, a differential capacitive \gls{dac}, a dynamic comparator, and asynchronous control logic~\cite{adc_brodersen}. 
Moreover, the control logic was implemented as dynamic to increase the speed of the ADC and, at the same time, to reduce power consumption.
Due to technological limitations in minimum capacitance available in the Process Design Kit (PDK), a split \gls{dac} architecture with split capacitor $C_s$ was used to reduce the \gls{dac} input capacitance. As a result, the effective unit capacitance is much lower than the minimum physical one used in the \gls{dac} design.
For additional power savings, all blocks were designed to dissipate power only during conversion, eliminating all static power.
Asynchronous logic was used to increase speed and eliminate the fast bit-cycling clock distribution, greatly simplifying the design of a multi-channel \gls{asic} and significantly improving the power budget. 

\begin{figure}[tbp]
 \centering
 \includegraphics[scale=0.35]{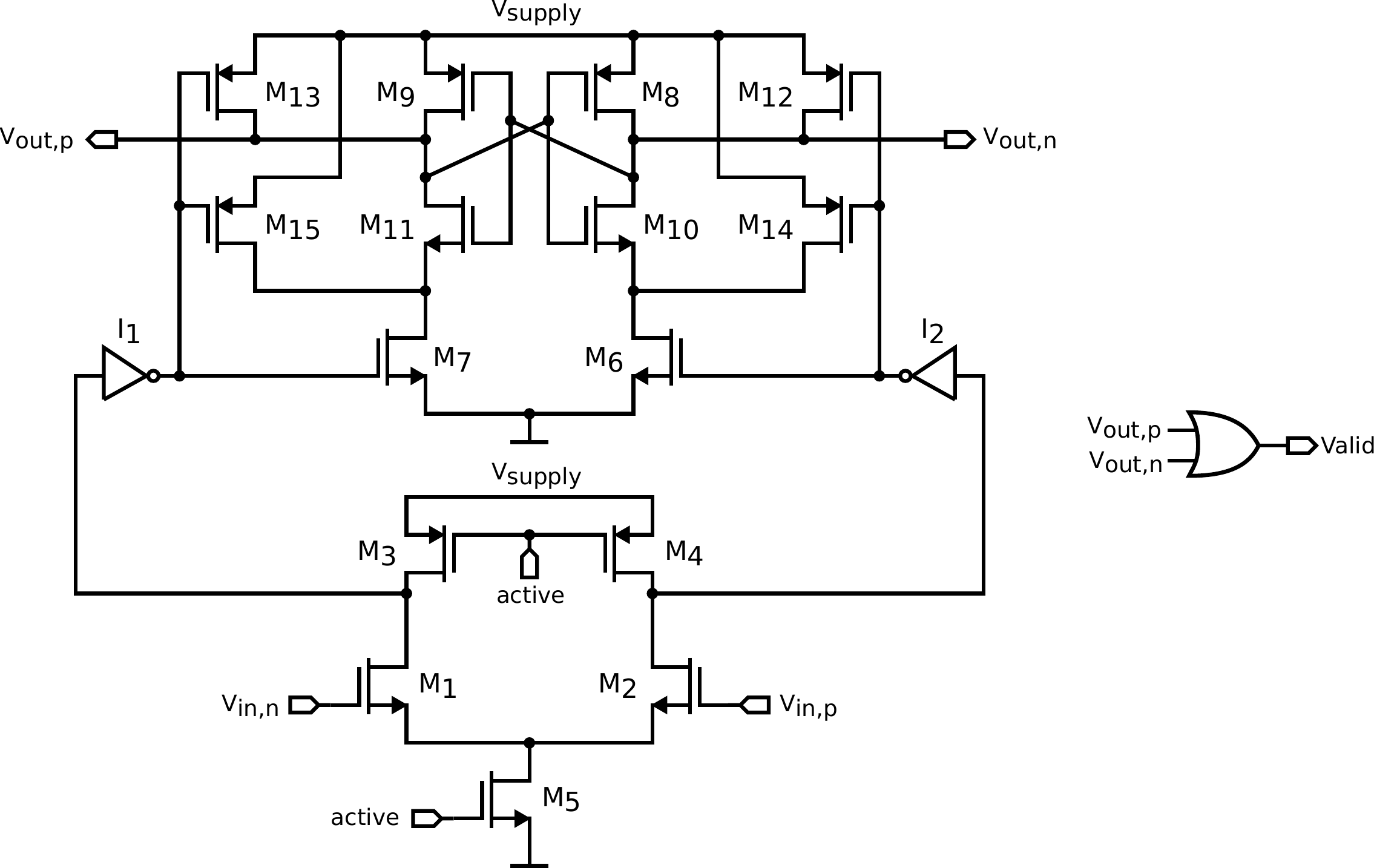}
 \caption{Schematic diagram of the dynamic comparator}
 \label{fig:sar_comparator}
\end{figure}

To explore and optimise \gls{adc} performance, several versions of \gls{sar}  \gls{adc} have been developed. The \glspl{adc} differ in the \gls{dac} switching scheme, the implementation of the \gls{dac} capacitors, and the power dissipated by the \gls{sar} logic. All \gls{adc} versions use the same bootstrapped sampling switch~\cite{adc_sumanen, bootstrap_switch} and dynamic comparator~\cite{comp_dynamic_jeon1}. 
The designed comparator, shown in Figure~\ref{fig:sar_comparator}, consists of two gain stages and an output latch. To symmetrise the circuit and minimise the effect of parasitics, a decision stage (generating the  \textit{Valid} signal) was added to the comparator core.

\subsection{Switching scheme and DAC}

\begin{figure}[tbp]
 \centering
 \includegraphics[width=0.98\textwidth]{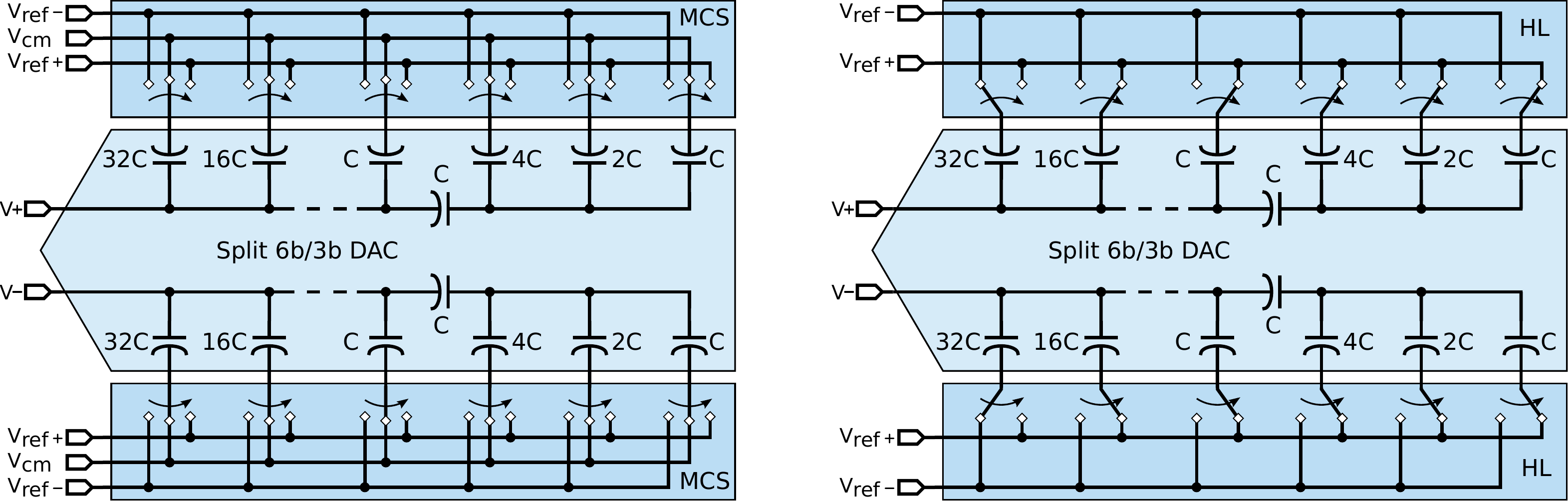}
 \caption{Block diagram of differential 9-bit \glspl{dac} with the MCS switching scheme (left) and the HL switching scheme (right).}
 \label{fig:sar_swschemes}
\end{figure}

Numerous \gls{dac} switching schemes have been proposed for \gls{sar} \gls{adc} to achieve the highest power efficiency~\cite{sar_survay}. In this work, two very efficient schemes have been used. The first one, the Merged Capacitor Switching (MCS) scheme~\cite{mcs,mcs1} shown in figure~\ref{fig:sar_swschemes}~(left), uses three reference voltages Vref+, Vref-, Vcm, but the accuracy of the \gls{dac} does not depend on the accuracy of the reference common voltage Vcm. Another great advantage of this scheme is that the \gls{dac} output common voltage is constant and equal to Vcm during conversion, which makes comparator operation easier. 
In the sampling phase, all \gls{dac} switches are connected to the Vcm voltage (as shown in figure~\ref{fig:sar_swschemes} (left)). 
After the first decision of the comparator, the upper/lower MSB switch (connected to 32~C) of \gls{dac} changes to Vref+/Vref- or Vref-/Vref+ depending on the result of the comparison. Such changes occur after each bit has been converted, so that at the end of the conversion there is no switch connected to Vcm.

The second switching scheme shown in figure~\ref{fig:sar_swschemes}~(right), called HL (HighLow) for brevity, is a modification of the scheme proposed by Sanyal and Sun~\cite{HLswitching}. The difference from this scheme is the use of Vref+, Vref- voltage references for all bits down to the least significant one, and omitting an additional common-mode reference Vcm. The consequence of not using Vcm is an additional capacitive branch of \gls{dac}, and thus doubling of the \gls{dac} capacitance. 
A disadvantage of this scheme is that the \gls{dac} output common mode voltage is not constant during conversion. 
In the sampling phase, all \gls{dac} switches, except the MSB bit, are connected to the Vref+ voltage, while the MSB switch is connected to Vref- (as shown in figure~\ref{fig:sar_swschemes}~(right)). 
After the first decision of the comparator, the upper or lower MSB switch (connected to 32~C) of \gls{dac} changes to Vref+ depending on the result of the comparison. After the conversion of each subsequent bit, the upper or lower \gls{dac} switch of the corresponding bit changes to Vref-, depending on the result of the comparison.

\begin{figure}[tbp]
 \centering
 \includegraphics[width=0.98\textwidth]{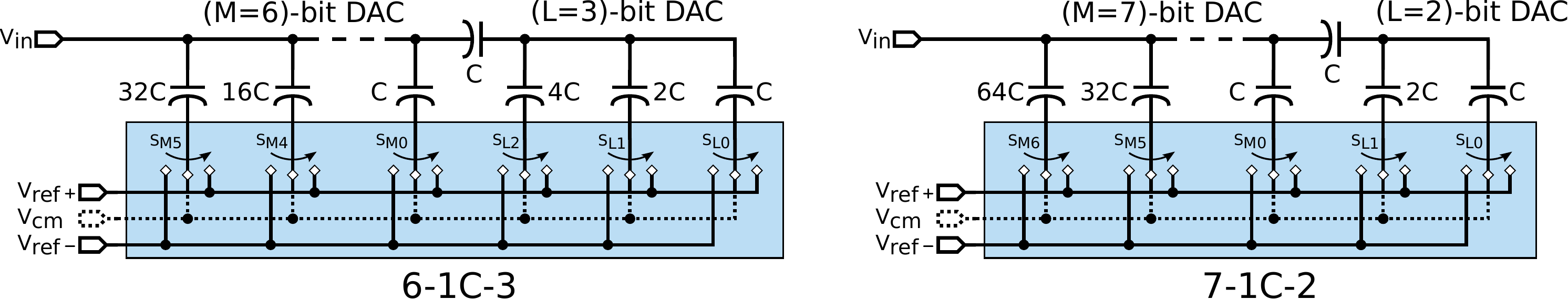}
 \caption{Comparison of DAC split used for MIM (left) and MOM (right) capacitors; for the sake of clarity only half of the DAC in MCS is presented (in HL the Vcm is omitted).}
 \label{fig:dac_splits}
\end{figure}

There are two types of capacitors available in the 65~nm CMOS process, MIM and MOM, so each version of \gls{dac} has been implemented using both of them. The MIM capacitor has a higher minimum value than the MOM, but its advantage is a lower parasitic capacitance of the top plate.
The MIM \gls{dac} was designed with a 6-bit sub-DAC for the most significant bits and a 3-bit sub-DAC for the least significant bits, as shown in figure~\ref{fig:dac_splits}~(left). 
The MOM \gls{dac}, for which a smaller minimum capacitance was available, was designed with a different split, using a 7-bit sub-DAC for the most significant bits (using 64 C in the MSB branch) and a 2-bit sub-DAC for the least significant bits (using 2 C in the MSB branch). The comparison of both \gls{dac} splits is shown in figure~\ref{fig:dac_splits}; for the sake of clarity, only half of the \gls{dac} is presented. 

\subsection{Asynchronous SAR logic}

The asynchronous design allows individual control of the time of subsequent conversion steps, providing the ability to find the best trade-off between effective \gls{adc} resolution and speed.  
In the first phase of the ADC operation, the sampling of the analogue input signal is performed in capacitive \gls{dac}. Then the main phase, analogue-to-digital conversion, is started. The sequence, which is repeated for each bit conversion, consists of the following steps: 
\begin{enumerate}
    \item initialisation of the comparator operation and waiting until its decision is made,
    \item memorizing the decision of the processed bit,
    \item toggling the switches setting the capacitive \gls{dac} voltage,
    \item waiting until the \gls{dac} voltage is settled precisely enough before the next bit conversion can start.
\end{enumerate}
During the last bit conversion, after step 2, the \gls{dac} is reset to the initial configuration and \gls{adc} is ready for the next conversion. Although steps 1 to 3 should be as fast as possible, the duration of step 4 can be optimised to find a compromise between a long enough settling time sufficient for the required \gls{adc} precision and the shortest possible time for the fastest \gls{adc} conversion. Since the required precision (and thus the settling time) is the highest for the MSB bit and the lowest for the LSB bit, the duration of step 4 can be optimised separately for different bits. 

\begin{figure}[tbp]
 \centering
 \includegraphics[width=0.8\textwidth]{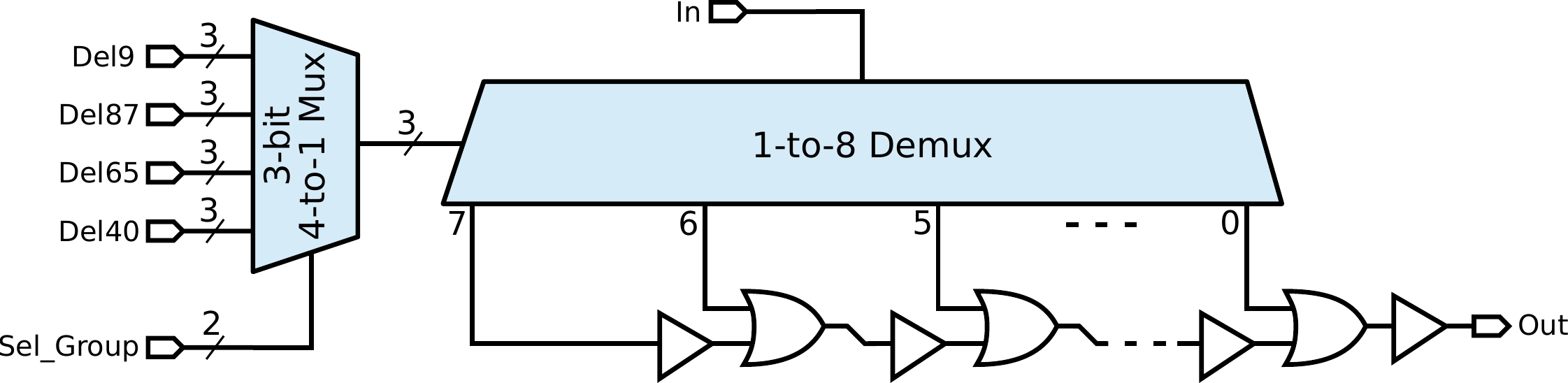}
 \caption{Simplified block diagram of the variable delay block.}
 \label{fig:delay_block}
\end{figure}

To achieve the best compromise between the highest effective resolution and the highest conversion rate, a variable delay has been introduced that adjusts the settling time of \gls{dac}. It may be optimised separately for different groups of bits, similar to what was done in~\cite{adc10b_130nm}.
The concept of this solution is shown in figure~\ref{fig:delay_block}.
There are four delays for four groups of bits: one for the most significant bit 9, denoted \textit{Del9}, the second for bits 8--7, denoted \textit{Del87}, the third for bits 6--5, denoted \textit{Del65}, and the last for the remaining bits 4--0, denoted \textit{Del40}. During conversion, the 2-bit \textit{Sel\_Group} signal selects the appropriate delay for the subsequent bit. Each of these delays can be set to eight different values (typically longer for more significant bits) controlled by a 3-bit register. The delays should be set experimentally to achieve the best performance of the given \gls{adc} prototype, and can be reused for all \glspl{adc} from the same production batch.

The \gls{sar} logic of the MCS and HL \glspl{adc} is very similar but adapted to correctly control the switches to Vref+, Vref-, Vcm in the MCS version and the switches to Vref+, Vref- in the HL version.
As the \gls{sar} logic is the largest contributor to the \gls{adc} power consumption, two versions of asynchronous control logic have been designed, focusing on the lowest power or the highest sampling rate. For the lowest power version, an additional condition was set to achieve a minimum sampling rate of 40~MSps. The logic implemented in both versions is functionally the same; the only difference being the transistor sizing.

\subsection{Design summary and layout}

In total, eight different \gls{sar} \gls{adc} versions were designed that differed in: \gls{dac} switching scheme (MCS, HL), \gls{dac} capacitor type (MIM, MOM), and standard or low-power (lp) consumption. They are denoted as follows: MCS-MIM, MCS-MIM-lp, MCS-MOM, MCS-MOM-lp, HL-MIM, HL-MIM-lp, HL-MOM, HL-MOM-lp.  

The prototype \glspl{adc} were fabricated in 65~nm CMOS technology, which has been proven to be a radiation hard technology. An additional basic precaution has been taken in the design of not using transistors of minimum size to improve \gls{adc} immunity to radiation damage.

One of the key design tasks was to draw a layout of capacitive \gls{dac} that would guarantee good \gls{adc} linearity. To achieve this, the guidelines listed below were followed.
\begin{enumerate}
    \item First, it is crucial to minimise parasitic capacitance seen from the LSB subDAC top plate (see figure~\ref{fig:sar_diagram}) to all constant potentials, as this parasitic directly degrades the \gls{dac} linearity.
    \item Next, you need to ensure that the values of all parasitic capacitances, parallel to subsequent capacitors in \gls{dac}, scale proportionally to them to maintain binary weighting.
    \item Finally, the parasitic capacitance seen from the top plate of the MSB subDAC to any constant potential should be reduced, but only in a way that does not affect the previous optimisations, since this parasitic only reduces the \gls{adc} input range, but has no effect on the \gls{dac} linearity.
    
\end{enumerate}

\begin{figure}[tbp]
 \centering
 \includegraphics[width=0.98\textwidth]{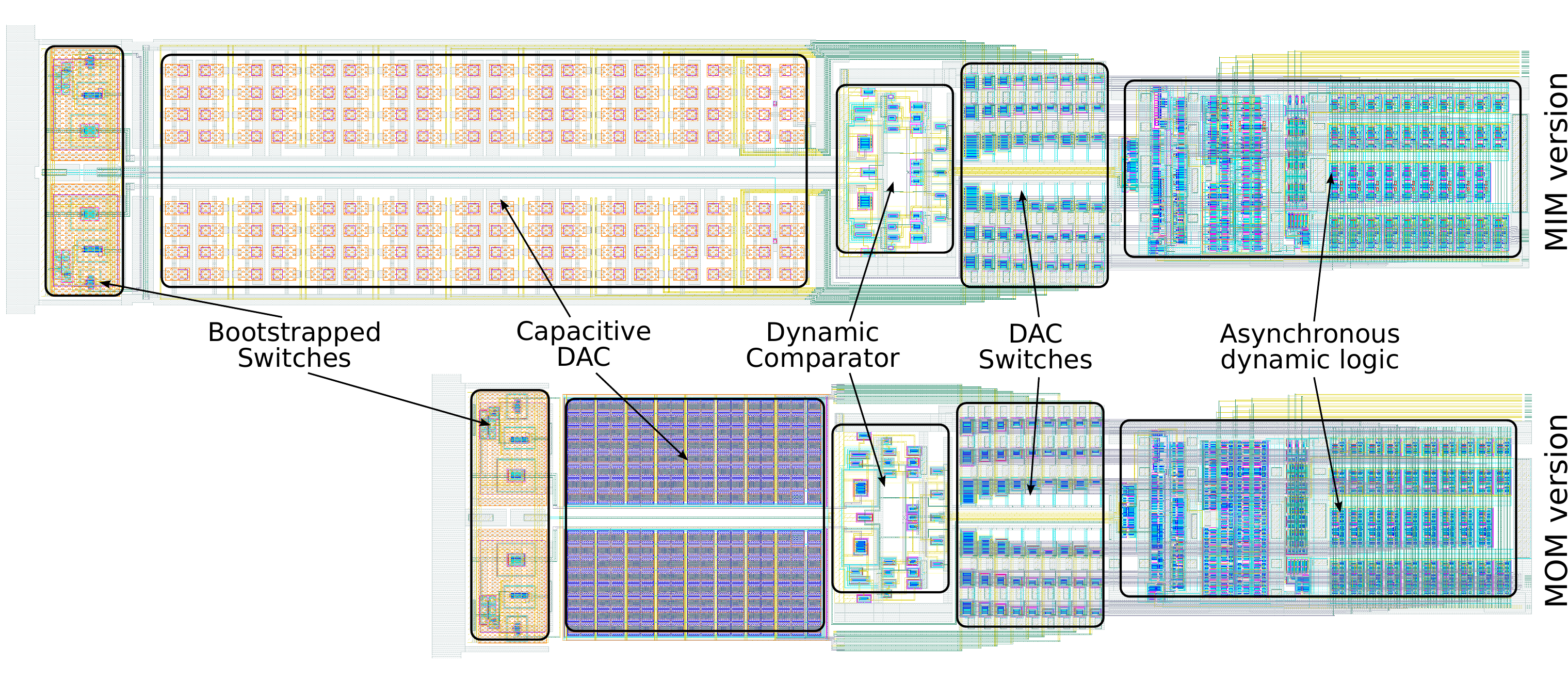}
 \caption{Layout of the \gls{adc} with MIM (top) and MOM (bottom) \gls{dac}.}
 \label{fig:sar_layout}
\end{figure}

The ADC layout was drawn in 60~$\upmu$m pitch to facilitate the implementation of a multi-channel readout \glspl{asic}.
The size of \glspl{adc} with MIM \gls{dac} is 330\,$\upmu$m $\times$ 60\,$\upmu$m and with MOM \gls{dac} 235\,$\upmu$m $\times$ 60\,$\upmu$m, as shown in figure~\ref{fig:sar_layout}.
The blocks from left to right are: bootstrap switches, capacitive \glspl{dac}, comparator, switches to reference voltages for subsequent \gls{dac} bits, and \gls{sar} control logic. 

\section{ADC measurements}

\begin{figure}[tbp]
 \centering
 \includegraphics[width=0.9\textwidth]{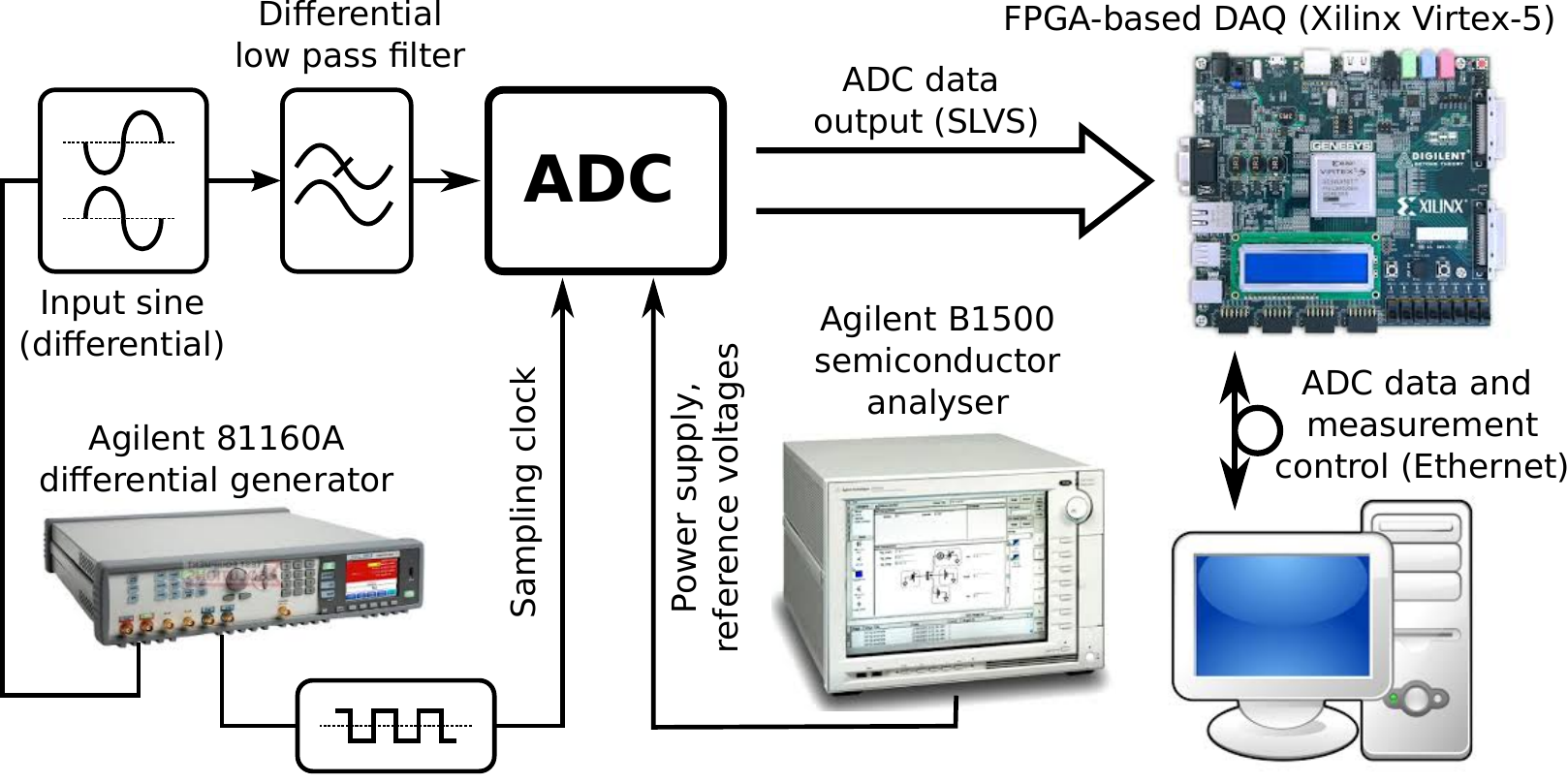}
 \caption{Setup for static and dynamic \gls{adc} measurements.}
 \label{fig:meas_setup}
\end{figure}

A dedicated FPGA-based setup was built as shown in figure~\ref{fig:meas_setup} to characterise the performance of \gls{adc} prototypes.
The Agilent B1500 semiconductor parameter analyser delivers the power supply, reference voltages, and measures the corresponding currents. It also generates input signals for static measurements. The Agilent 81160A generates the sampling clock and input signals for dynamic measurements. The data acquisition system was built based on the Xilinx Virtex-5 FPGA incorporated into the Genesys evaluation board.

Using this setup, standard static measurements, that is, \gls{inl} and \gls{dnl}, as well as basic dynamic metrics, that is, \gls{snhr}, \gls{thd}, \gls{sfdr}, \gls{sinad}, and \gls{enob}, were obtained. Due to the limitations of the setup, the dynamic metrics were measured at the 0.1 Nyquist input frequency, so the effective number of bits at lower input frequencies, called $\enoblf$ was calculated.

\subsection{Internal delay optimisation and static measurements}

 In the first series of measurements, the static \gls{dnl} and \gls{inl} errors were measured at a sampling frequency of 10~MHz for different settings of \gls{adc} internal delays.  The delays for each group of bits were tuned to achieve the best \gls{adc} performance, with the goal of eliminating missing codes (not always possible) and obtaining the best compromise between acceptable \gls{dnl}, \gls{inl} errors (it was not always possible to get it below $1$ LSB) and the shortest possible delays.

After optimisation, performed separately for each \gls{adc} version, the internal delays were set as shown in table~\ref{tab_delay}.
These settings were used for all the following measurements.
It should be noticed that unit delays of standard and low power versions are different, so the delay settings cannot be compared directly.
\begin{table*}[htb!]
\caption{Optimised internal delay settings for all \gls{adc} versions.}
\label{tab_delay}
\begin{center}
\begin{tabular}{| c | c | c | c | c|}
\hline
& Del9 & Del87 & Del65 & Del40 \\
\hline \hline
MCS-MIM & 2 & 0 & 0  & 0 \\
\hline
MCS-MOM & 3 & 2 & 1  & 1 \\
\hline
HL-MIM & 2 & 1 & 1  & 0 \\
\hline
HL-MOM & 1 & 1 & 1  & 1 \\
\hline
\hline
MCS-MIM-lp & 3 & 2 & 1  & 1 \\
\hline
MCS-MOM-lp & 3 & 2 & 1  & 1 \\
\hline
HL-MIM-lp & 7 & 1 & 1  & 0  \\
\hline
HL-MOM-lp & 4 & 2 & 0  & 0  \\
\hline
\end{tabular}
\end{center}
\end{table*}

\begin{figure}[ht!]
\centering
 \includegraphics[width=0.98\columnwidth]{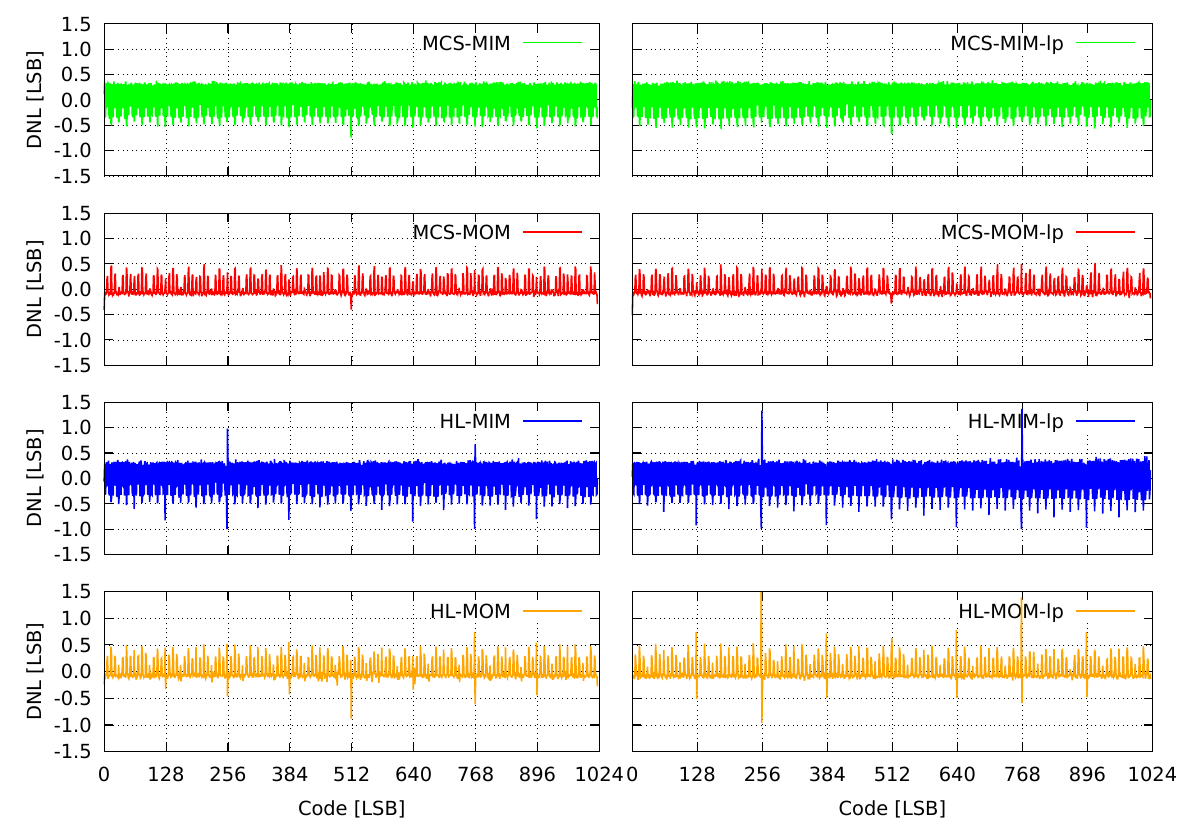}
 \caption{DNL error for all \gls{adc} versions measured at 10~MHz sampling frequency.}
 \label{fig:static_dnl}
\end{figure}

\begin{figure}[ht!]
 \centering
 \includegraphics[width=0.98\columnwidth]{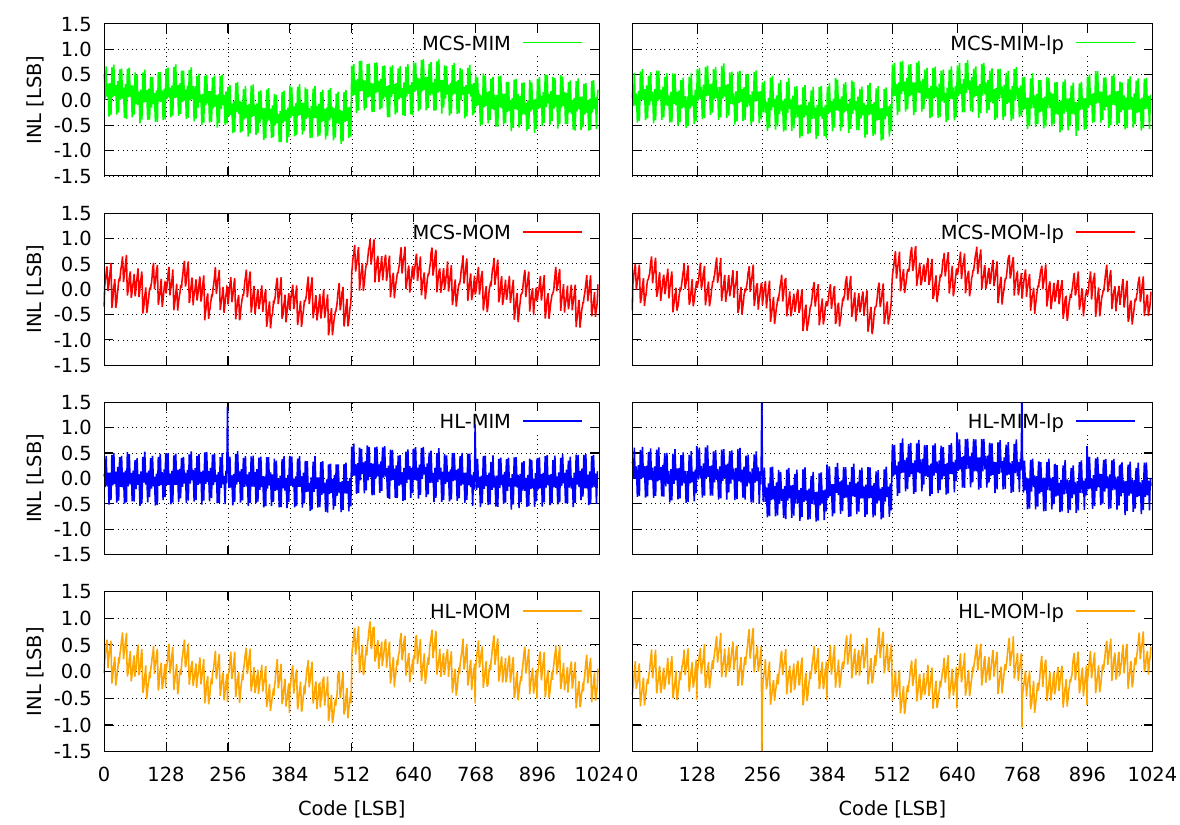}
 \caption{INL error for all \gls{adc} versions measured at 10~MHz sampling frequency.}
 \label{fig:static_inl}
\end{figure}

The results of \gls{dnl} and \gls{inl} errors obtained for all \gls{adc} versions are shown in figures~\ref{fig:static_dnl} and~\ref{fig:static_inl} respectively.

Analysing the results, one can observe several features. 

\begin{itemize}
    \item Analysis of table \ref{tab_delay} allows to formulate some tentative conclusions.  The MCS-MIM configuration is expected to be the fastest as its total delay is the smallest. Furthermore, there is potential room for future improvements in ADC performance by shortening delays because some less significant bits (mainly \textit{Del65} and \textit{Del40}) have delays set to zero. On the other hand, \textit{Del9} for the HL-MIM-lp version is 7, suggesting that the maximum delay is too small and results in many missing codes (\gls{dnl}$<-0.9$) as seen in figure~\ref{fig:static_dnl}.  
    
    \item \glspl{adc} with the HL switching scheme show worse linearity (spikes in \gls{inl} and \gls{dnl}) than the MCS versions, for which both \gls{inl} and \gls{dnl} errors show good performance remaining always below 1~LSB. For \glspl{adc} with the HL switching scheme, except for HL-MOM, there are always one or two missing codes and one or two codes with absolute INL or DNL errors greater than one. The HL-MIM-lp version typically shows six missing codes, and for this reason this version was not used in further studies. Since in both MCS and HL versions the same \glspl{dac} were used, the worst non-linearity errors are not caused by the \gls{dac} layout imperfections or parasitics. A possible reason may be poorer performance of the comparator, particularly during the MSB conversion, which does not work at constant common mode voltage in the HL switching scheme.
    
    \item Comparing the INL results, it is seen that both the MIM and MOM \gls{dac} versions show similar behaviour (except HL-MOM-lp) and have a large change of \gls{inl} at the centre code (512). This is probably due to the layout imperfections of the differential \gls{dac}.
    
    \item The standard and low power \gls{adc} versions have similar INL and DNL behaviour (except for the HL-MOM and HL-MOM-lp versions) and also do not differ much quantitatively. This could be expected since they differ only in the dimensioning of the transistors in \gls{sar} logic.
\end{itemize}

\subsection{Dynamic parameter measurements}

\begin{figure}[ht!]
\centering
 \includegraphics[width=0.98\columnwidth]{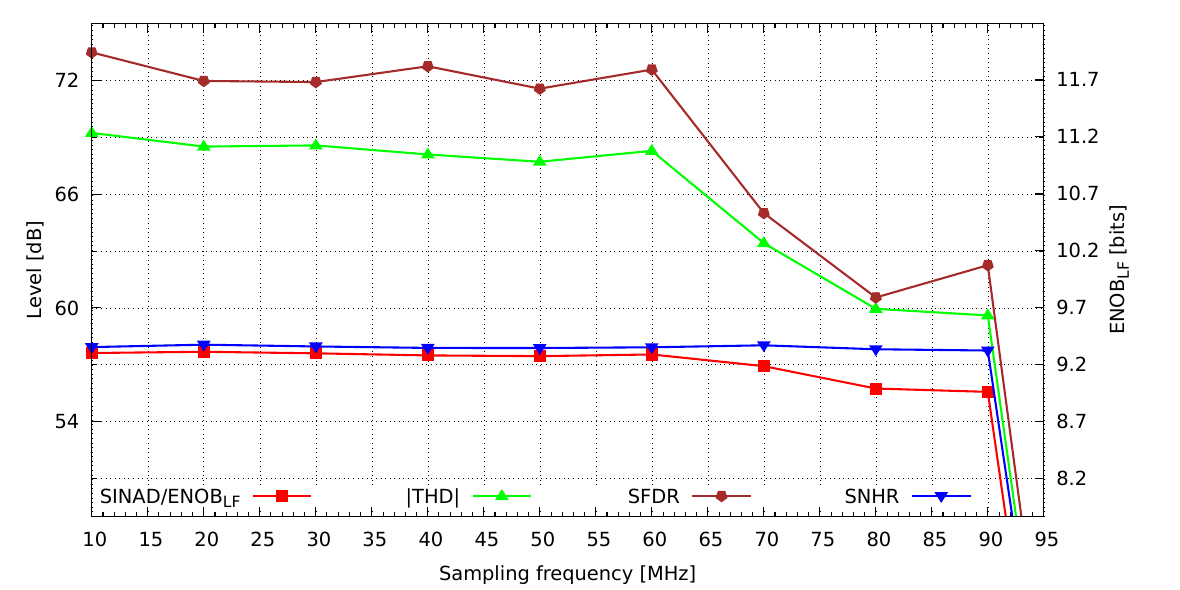}
 \caption{Measurement of dynamic \gls{adc} metrics in function of sampling frequency at 0.1 Nyquist input signal frequency, for the MCS-MIM \gls{adc} version.}
 \label{fig:dynamic_example}
\end{figure}

Dynamic \gls{adc} metrics were measured for all \gls{adc} versions as a function of the sampling frequency at 0.1 Nyquist input signal frequency. The main dynamic parameters \gls{sinad}, \gls{thd}, \gls{snhr}, \gls{sfdr}, and $\enoblf$ are presented in figure~\ref{fig:dynamic_example} for the MCS-MIM \gls{adc} version.
It is seen that $\enoblf$ is saturated at about 9.3 for lower sampling frequencies, starts to decrease above 60~MHz reaching $\sim$8.9 at 90~MHz, and decreases sharply for higher sampling frequencies.

\begin{figure}[ht!]
\centering
 \includegraphics[width=0.98\columnwidth]{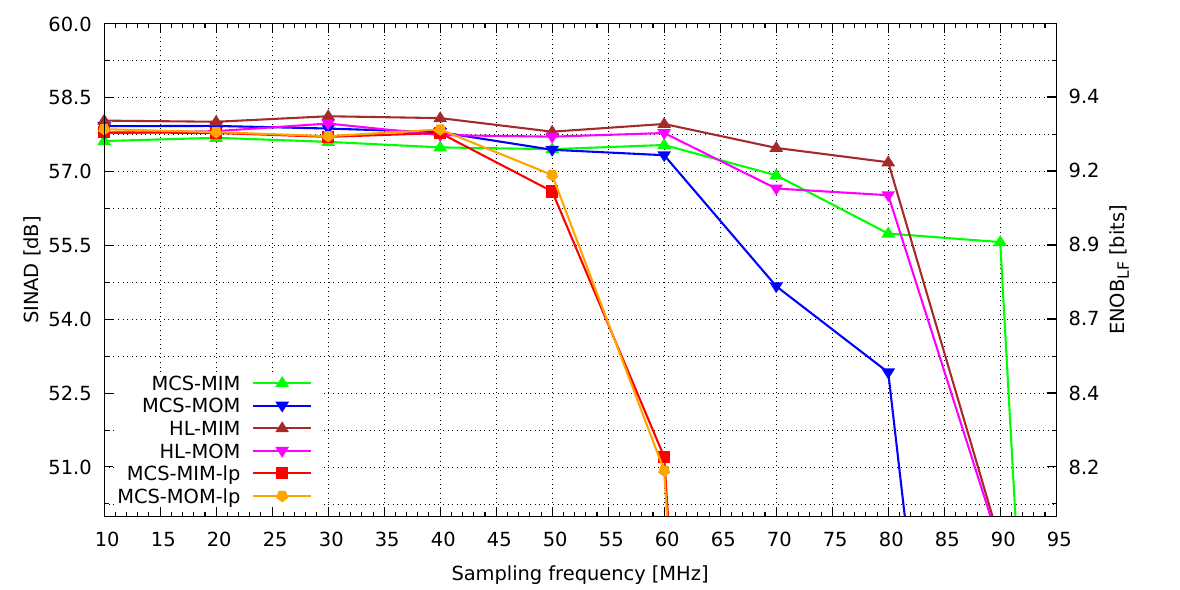}
 \caption{Comparison of SINAD and ENOB as a function of sampling frequency for different ADC versions.}
 \label{fig:dynamic_compare}
\end{figure}

The comparison of SINAD and ENOB for all \gls{adc} prototypes (except HL-MIM-lp and HL-MOM-lp) is shown in figure~\ref{fig:dynamic_compare}. The HL-MOM-lp \gls{adc} version is not shown and will not be used in future studies because its effective resolution is 1--2 bits worse than for other versions shown. 
The HL-MIM version of the ADC, although it has worse non-linearity errors, shows $\enoblf$ like the other versions. The two codes with worse \gls{inl} and \gls{dnl} errors do not significantly affect $\enoblf$ for this version.

As expected, the standard versions of \gls{adc} are much faster than the low-power versions. In fact, MCS-MIM has an $\enoblf$ of 8.9 bits up to 90~MHz, while the HL-MIM and HL-MOM versions have an $\enoblf$ of more than 9 bits up to 80~MHz. The best versions of the low-power \glspl{adc}, MCS-MOM-lp and MCS-MIM-lp keep $\enoblf$ above 9 bits up to 50~MHz.


\subsection{Power consumption}
\begin{figure}[ht!]
\centering

 \includegraphics[width=0.98\columnwidth]{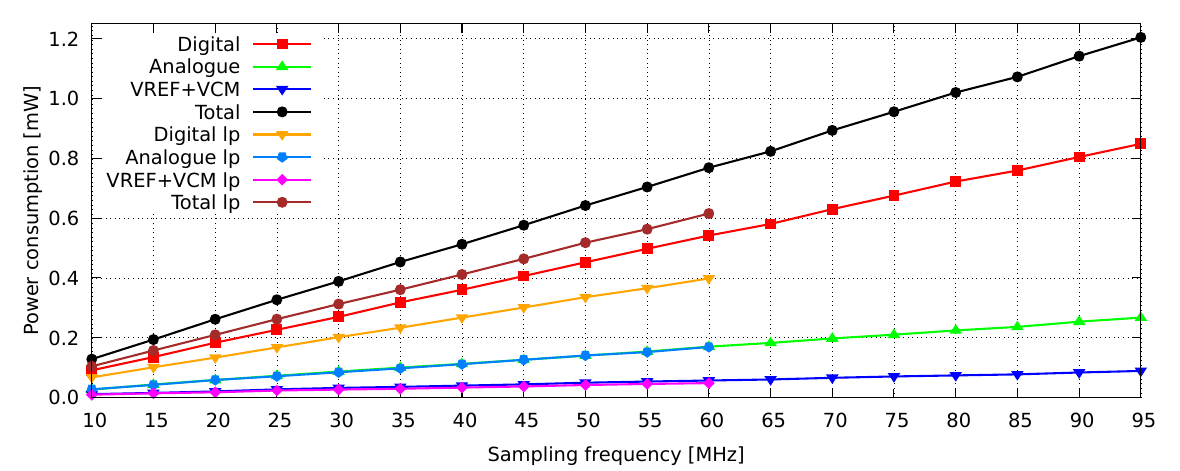}
 \caption{Contributions to power consumption as a function of sampling frequency for the standard MCS-MIM and the low-power MCS-MIM-lp \gls{adc} versions.}
 \label{fig:power_example}
\end{figure}

Power consumption was measured for different \gls{adc} versions as a function of the sampling frequency, up to the frequencies at which the ADC performed well (lower for low-power versions of the ADC).
Contributions to the total power of key \gls{adc} blocks are presented in figure~\ref{fig:power_example} for the MCS-MIM and MCS-MIM-lp ADC versions. 
As expected, the power consumption is proportional to the sampling rate, and the different power contributions of the MCS-MIM and MCS-MIM-lp versions overlap, except for the digital part (and so the total power). Total power is extremely low, reaching 1~mW at about 80~MHz sampling frequency for MCS-MIM \gls{adc}.
About two-thirds of the total power comes from the digital \gls{adc} part, while less than a third comes from the analogue part (comparator and bootstrapped switches), and the smallest contribution comes from the reference voltages.

\begin{figure}[ht!]
\centering
 \includegraphics[width=0.98\columnwidth]{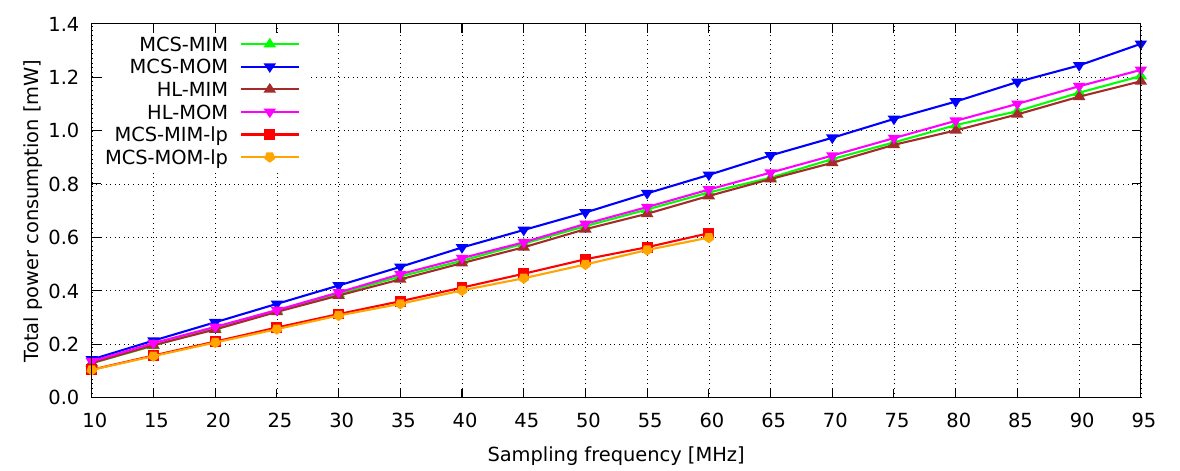}
 \caption{Comparison of total power consumption as a function of sampling frequency for different \gls{adc} versions.}
 \label{fig:power_compare}
\end{figure}

The comparison of the total power for different \glspl{adc} is shown in figure~\ref{fig:power_compare}.
Two groups of straight curves are clearly visible, the curves for standard and low-power \glspl{adc}. The power consumption of standard \gls{adc} versions is slightly above 500~$\upmu$W at 40~MHz sampling frequency and slightly above 1~mW at 80~MHz. The low-power \glspl{adc} are slower and work well up to about 55~MHz sampling frequency, but their power consumption at 40~MHz is only about 400~$\upmu$W, more than 20\% less than the standard \glspl{adc}.


\subsection{The \gls{adc} figure of merit}

Using the effective \gls{adc} resolution and the power consumption, the well-known Walden \gls{adc} \gls{fom}~\cite{walden2} can be calculated:
\begin{equation}
        \fom = \frac{\mathrm{Power}}{2^{\mathrm{ENOB}} \cdot f_{\mathrm{sample}}},
\end{equation}
where \gls{enob} is typically measured at the Nyquist frequency of the input signal. 
The low-frequency Figure of Merit ($\fomlf$) is also often used, which is calculated using the effective resolution $\enoblf$ obtained at the lower frequency of the input signal.

\begin{figure}[ht!]
\centering
\includegraphics[width=0.98\columnwidth]{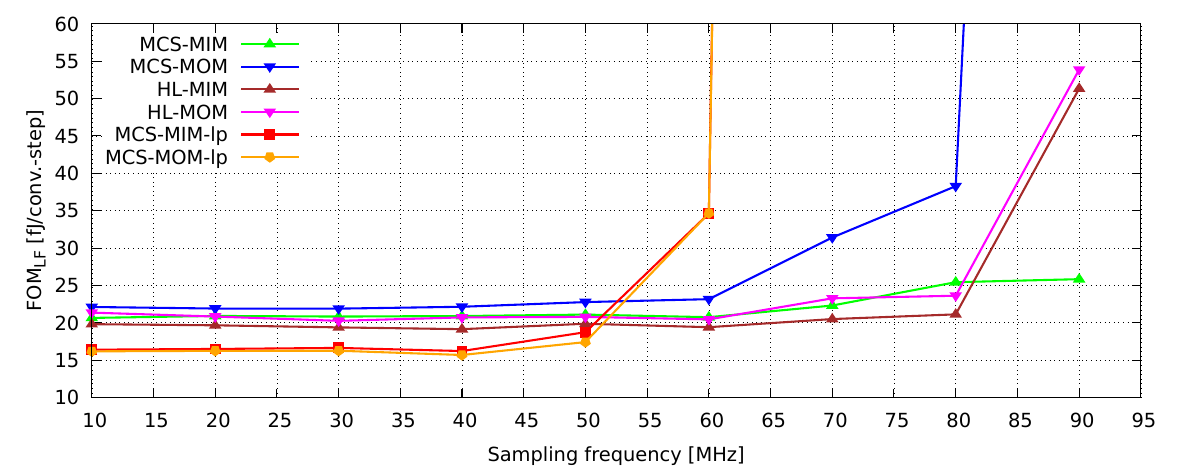}
 \caption{Comparison of $\fomlf$ as a function of sampling frequency for different \gls{adc} versions.}
 \label{fig:fom_all}
\end{figure}

The comparison of $\fomlf$ for the six best \gls{adc} versions (except HL-MIM-lp and HL-MOM-lp) is presented in figure~\ref{fig:fom_all}.
The standard \glspl{adc} are characterised by $\fomlf$ between 20--26\,\fomunit{} over most of the usable sampling frequency range, while the low power \glspl{adc} have, as expected, an even better $\fomlf$, in the range 15--20\,\fomunit{}.

\section{Comparison to the state-of-the-art}

In table~\ref{tab_comp} the key parameters of the standard (HL-MOM) and low-power (MCS-MOM-lp) \gls{adc} versions are compared with the state-of-the-art \glspl{adc} with the same resolution and similar sampling rates (but at least 40~MSps), designed in similar size CMOS technologies.
\begin{table*}[htb!]
\caption{Comparison with the state-of-the-art \glspl{adc}.}
\label{tab_comp}
\begin{center}
\begin{tabular}{| c | c | c | c | c | c | c | c|}
\hline
&  &  &  &  &  & MCS- & HL-\\
& \cite{adc40nm10b120M} & \cite{adc40nm10b100M} & \cite{sar_adc_5} & \cite{adc65nm10b50M} & \cite{adc65nm10b90M} & MOM-lp & MOM \\
\hline \hline
Architecture & SAR & SAR  & SAR & SAR & SAR  & SAR & SAR \\
\hline
CMOS [nm] & 40 & 40 & 90 & 65 & 65 & 65 & 65 \\
\hline
Resolution [bits] & 10 &  10  & 10 & 10 & 10  & 10 & 10 \\
\hline
Supply [V] & 1.2  & 1.1/1.3 & 1.2 & 1 & 1  & 1.2 & 1.2 \\
\hline
Area [mm$^2$] & 0.023 & 0.027 & 0.024 & 0.039 & 0.135 & 0.014 & 0.014 \\
\hline
$C_{in}$ [pF] (each input) & 1 & 1 & 1.78 & 0.51 & 1.9 & 0.6 & 0.6 \\
\hline
$f_\mathrm{sample}$ [MHz] & 120 & 100 & 50 & 50 & 90 & 40 & 80 \\
\hline
Power [$\upmu$W]$^{\footnotesize{\mathrm{a}}}$ & 1120 & 1090  & 664 & 820 & 1760 & 402 & 1040 \\
\hline
Max INL [LSB] & 0.6 & 0.82  & 0.45 & <0.82 & - & 0.89 & 0.97 \\
\hline
Max DNL [LSB] & 0.73 & 0.73  & 0.36 & <0.72 & - & 0.52 & 0.88 \\
\hline
$\enoblf$ [bits] & 9.26 & 9.3  & 9.26 & 9.16 & 8.8 & 9.32 & 9.1 \\
\hline
$\fomlf$ [\fomunit{}] & 15.2 & 17.3 & 21.68 & 28.7 & 44  & 15.69 & 23.61 \\
\hline
ENOB [bits] & 8.83 & 9.06 & - & 9.1 & 8.7 & - & - \\
\hline
$\fom$ [\fomunit{}] & 20.5 & 20.4 & - & 29.7 & 47 &  - & - \\
\hline
\multicolumn{8}{l}{${}^{\footnotesize{\mathrm{a}}}$\footnotesize{Power consumption of the reference voltage is not included (either external or internal). }}\\
\end{tabular}
\end{center}
\end{table*}

The $\fomlf$ of the state-of-the-art \glspl{adc} is between 15--30\,\fomunit{} and the designs presented in this work also stay well in this range.
This is due to the fact that the most important parameters, such as power/frequency or effective resolution, obtained in this work are similar to the state-of-the-art \glspl{adc}. Furthermore, the \gls{adc} size (plus small pitch) of this work compares very well with other designs, which is particularly important considering applications in multi-channel \glspl{asic}. 
Since all the above designs strive to achieve maximum speed with minimum power and minimum area, this comes at the cost of the resulting ENOB, which is noticeably lower (at least 0.7~LSB) than the nominal (10 bits).


\section{Conclusion}

The design and measurements of a fast ultra-low power 10-bit \gls{sar} \glspl{adc} in CMOS 65~nm process have been presented.
The measurements performed confirm very good \gls{adc} functionality, reflected in $\enoblf$ of about 8.9--9.1 bits up to maximum sampling frequencies of 80--90\,MHz, ultra-low power of about 1~mW at 80\,MHz, and excellent $\fomlf$ of 24--26\,\fomunit{} at 80\,MHz. The low-power \gls{adc} versions work well up to about 50\,MHz sampling frequency, achieving an $\enoblf$ of about 9.3 bits at 40\,MHz with a power consumption of about 400~$\upmu$W, corresponding to very low $\fomlf$ below 16\,\fomunit{}. Measurements have shown that the prototype \glspl{adc} are fully functional with both the MCS and HL switching schemes, although the MCS version is more robust to non-linearity errors.
The designed \glspl{adc} are ready for implementation in multi-channel readout \glspl{asic}.

One of the \glspl{adc} (MCS-MIM version) has already been implemented in the monitoring subsystem of the Low Power Giga Bit Transceiver (lpGBT) \gls{asic}~\cite{lpGBT_mon}, the common serialiser/deserialiser device for the Large Hadron Collider (LHC) detectors. As this application does not require high sampling rates, while the design was added directly to the production version of the lpGBT before the \gls{adc} prototype was available and verified, in the actual implementation, for safety, the asynchronous \gls{sar} logic was replaced by an automatically synthesised synchronous control logic. The lpGBT tests confirmed the very good performance of the \gls{adc} and showed that it works correctly during irradiation up to at least 3.8~MGy dose~\cite{lpGBT_mon}.

\acknowledgments
This work has received funding from the Polish Ministry of Science and Higher Education under contract No 5179/H2020/2021/2, and from the European Union’s Horizon 2020 Research and Innovation programme under grant agreement No 101004761 (AIDAinnova).


\end{document}